\documentclass[10pt, conference]{IEEEtran}

\usepackage[table]{xcolor}
\usepackage{colortbl}
\usepackage{mathtools}

\usepackage[english]{babel}
\usepackage{cite}
\usepackage{color}
\usepackage{xcolor} 
\usepackage{booktabs}
\usepackage{csquotes}
\usepackage{tablefootnote}
\usepackage{tabularray}
\usepackage{numprint}
\usepackage{nicematrix}
\usepackage[compat=0.6]{yquant}
\usepackage{lipsum}
\usepackage{comment}
\usepackage{inputenc}
\usepackage{fontenc}

\usepackage{algorithm}
\usepackage{algpseudocode}

\usepackage{subcaption}

\usepackage{pgfplots}
\pgfplotsset{compat = newest}
\usepackage{tikz}
\usetikzlibrary{arrows.meta, shapes.geometric,
                positioning,fit,shadows.blur,
                patterns,calc, matrix,
                tikzmark, backgrounds}
\usepackage{fp}
\usepackage{xspace}
\usepackage{amsmath,amssymb,amsfonts}
\usepackage{amsthm} 
\usepackage{bbm}
\usepackage{quantikz}
\usepackage{placeins}
\usepackage[T1]{fontenc}
\usepackage[colorlinks=true,urlcolor=teal,
            linkcolor=teal,citecolor=teal]{hyperref}
\usepackage{orcidlink}
\usepackage[most]{tcolorbox}
\usepackage{listings}
\usepackage{censor}
\usepackage{todonotes}
\usepackage[printonlyused, nolist]{acronym}
\usepackage[bold,full]{complexity}
\definecolor{lfdblack}{HTML}{000000}
\definecolor{lfdyellow}{HTML}{E69F00}
\definecolor{lfddgrey}{HTML}{999999}
\definecolor{lfdgreen}{HTML}{009371}
\definecolor{lfdhgrey}{HTML}{beaed4}
\definecolor{lfdred}{HTML}{ed665a}
\definecolor{lfdblue}{HTML}{1f78b4}
\definecolor{bggray}{gray}{0.9}

\newboolean{anonymous}
\setboolean{anonymous}{false}

\addto\extrasenglish{%
}

\newcommand{\reprolink}{https://github.com/lfd/QCE24-IndustryQubo}
\newcommand{\ie}{\emph{i.e.}\xspace}
\newcommand{\eg}{\emph{e.g.}\xspace}
\newcommand{\etal}{\emph{et al.}\xspace}

\newcommand{\repro}{\href{\reprolink}{reproduction package}\xspace}

\makeatletter
\def\calcLength(#1,#2)#3{%
  \pgfpointdiff{\pgfpointanchor{#1}{center}}%
  {\pgfpointanchor{#2}{center}}%
  \pgf@xa=\pgf@x%
  \pgf@ya=\pgf@y%
  \FPeval\@temp@a{\pgfmath@tonumber{\pgf@xa}}%
  \FPeval\@temp@b{\pgfmath@tonumber{\pgf@ya}}%
  \FPeval\@temp@sum{(\@temp@a*\@temp@a+\@temp@b*\@temp@b)}%
  \FProot{\FPMathLen}{\@temp@sum}{2}%
  \FPround\FPMathLen\FPMathLen5\relax
  \global\expandafter\edef\csname #3\endcsname{\FPMathLen}
}
\makeatother

\makeatletter
\newcommand{\linebreakand}{%
  \end{@IEEEauthorhalign}
  \hfill\mbox{}\par
  \mbox{}\hfill\begin{@IEEEauthorhalign}
}
\makeatother

\lstdefinestyle{query}{
  language=SQL,
  stepnumber=1,
  numbersep=10pt,
  tabsize=4,
  showspaces=false,
  showstringspaces=false,
  basicstyle=\linespread{1}\fontfamily{lmtt}\selectfont\small,
  keywordstyle=\color{blue},
  stringstyle=\color{purple},
  upquote=true,
  breaklines=true,
  commentstyle=\color{CadetBlue}
}

\definecolor{mygray}{rgb}{0.643,0.643,0.643}
\newtcolorbox{querybox}[2][]{%
  sidebyside align=top,
  enhanced,
  boxsep=2pt,
  arc=0pt,
  top=-3pt, bottom=-3pt,
  left=2pt, right=0pt,
  colback=white,
  colframe=mygray,
  boxrule=0.5pt,
  leftrule=12pt,
  overlay unbroken and first ={%
      \node[rotate=90,
        minimum width=0.5cm,
        anchor=south,
        yshift=-11pt,
        white]
      at (frame.west) {#2};
    }
}

\newtcolorbox{matrixbox}[2][]{%
  sidebyside align=top,
  enhanced,
  boxsep=0pt,
  arc=0pt,
  left=-1em,
  top=-0.8em,
  boxrule=0pt,
  colframe=bggray,
  colback=bggray,
  leftrule=12pt,
  overlay unbroken and first ={%
      \node[rotate=90,
        minimum width=0.5cm,
        anchor=south west,
        font=\itshape,
        yshift=0pt,
        xshift=0.5em,
        black]
      at (frame.south west) {#2};
    }
}

\ifbool{anonymous}{}{\renewcommand{\censor}[1]{#1}}
\ifbool{anonymous}{}{\renewcommand{\blackout}[1]{#1}}
\ifbool{anonymous}{\newcommand{\genemail}[2]{\href{mailto:xxx.xxx@xx.xx}{\blackout{#2}}}}{\newcommand{\genemail}[2]{\href{#1}{#2}}}

\begin{document}
\bstctlcite{BSTcontrol}

\begin{acronym}[qaoa]
  \acro{qaoa}[QAOA]{Quantum Approximate Optimization Algorithm}
  \acro{lrqaoa}[LR-QAOA]{Linear Ramp Quantum Approximate Optimization Algorithm}
  \acro{qubo}[QUBO]{Quadratic Unconstrained Binary Optimization}
  \acro{pubo}[PUBO]{Polynomial Unconstrained Binary Optimization}
  \acro{pbf}[PBF]{Pseudo-Boolean Function}
  \acro{op}[OP]{Optimization Problem}
  \acro{cop}[COP]{Combinatorial Optimization Problem}
  \acro{nisq}[NISQ]{Noisy Intermediate Scale Quantum}
  \acro{qecc}[QECC]{Quantum Error Correcting Code}
\end{acronym}

\title{Path Matters: Industrial Data\\ Meet Quantum Optimization}
\author{
\IEEEauthorblockN{\blackout{Lukas Schmidbauer\orcidlink{0009-0001-7171-0865}}}
  \IEEEauthorblockA{\blackout{\textit{Technical University of}}\\
    \blackout{\textit{Applied Sciences Regensburg}} \\
    \blackout{Regensburg, Germany} \\
    \genemail{mailto:lukas.schmidbauer@othr.de}{lukas.schmidbauer@othr.de}}
\and
  \IEEEauthorblockN{\blackout{Carlos A.\ Riofrío\orcidlink{0000-0002-7346-9198}}}
  \IEEEauthorblockA{
    \blackout{\textit{BMW AG}}\\
    \blackout{Munich, Germany}\\
    \genemail{mailto:carlos.riofrio@bmwgroup.com}{carlos.riofrio@bmwgroup.com}}
\and
    \IEEEauthorblockN{\blackout{Florian Heinrich}}
  \IEEEauthorblockA{
    \blackout{\textit{BMW AG}}\\
    \blackout{Munich, Germany}\\
    \genemail{mailto:florian.heinrich@bmw.de}{florian.heinrich@bmw.de}}
\and
  \IEEEauthorblockN{\blackout{Vanessa Junk\orcidlink{0000-0003-1675-8548}}}
  \IEEEauthorblockA{
    \blackout{\textit{OptWare GmbH}}\\
    \blackout{Regensburg, Germany}\\
    \genemail{mailto:vanessa.junk@optware.de}{vanessa.junk@optware.de}}
\linebreakand
  \IEEEauthorblockN{\blackout{Ulrich Schwenk}}
  \IEEEauthorblockA{
    \blackout{\textit{OptWare GmbH}}\\
    \blackout{Regensburg, Germany}\\
    \genemail{mailto:ulrich.schwenk@optware.de}{ulrich.schwenk@optware.de}}
\and
  \IEEEauthorblockN{\blackout{Thomas Husslein\orcidlink{0009-0008-7647-9376}}}
  \IEEEauthorblockA{
    \blackout{\textit{OptWare GmbH}}\\
    \blackout{Regensburg, Germany}\\
    \genemail{mailto:thomas.husslein@optware.de}{thomas.husslein@optware.de}}
\and
  \IEEEauthorblockN{\blackout{Wolfgang Mauerer\orcidlink{0000-0002-9765-8313}}}
  \IEEEauthorblockA{\blackout{\textit{Technical University of}}\\
    \blackout{\textit{Applied Sciences Regensburg}}\\
    \blackout{\textit{Siemens AG, Technology}}\\
    \blackout{Regensburg/Munich, Germany}\\
    \genemail{mailto:wolfgang.mauerer@othr.de}{wolfgang.mauerer@othr.de}}
}

\maketitle

\begin{abstract}
Real-world optimization problems must undergo a series of transformations before becoming solvable on current quantum hardware. 
Even for a fixed problem, the number of possible transformation paths—from industry-relevant formulations through binary constrained linear programs (BILPs), to quadratic unconstrained binary optimization (QUBO), and finally to a hardware-executable representation—is remarkably large. 
Each step introduces free parameters, such as Lagrange multipliers, encoding strategies, slack variables, rounding schemes or algorithmic choices---making brute-force exploration of all paths intractable. 
In this work, we benchmark a representative subset of these transformation paths using a real-world industrial production planning problem with industry data: the optimization of work allocation in a press shop producing vehicle parts. 
We focus on QUBO reformulations and algorithmic parameters for both quantum annealing (QA) and the Linear Ramp Quantum Approximate Optimization Algorithm (LR-QAOA). 
Our goal is to identify a reduced set of effective configurations applicable to similar industrial settings. 
Our results show that QA on D-Wave hardware consistently produces near-optimal solutions, whereas LR-QAOA on IBM quantum devices struggles to reach comparable performance. 
Hence, the choice of hardware and solver strategy significantly impacts performance. 
The problem formulation and especially the penalization strategy determine the solution quality. 
Most importantly, mathematically-defined penalization strategies are equally successful as hand-picked penalty factors, paving the way for automated QUBO formulation. 
Moreover, we observe a strong correlation between simulated and quantum annealing performance metrics, offering a scalable proxy for predicting QA behavior on larger problem instances.

\end{abstract}


\section{Introduction}
\label{sec:intro}

Optimization problems are widespread in industrial processes. Great effort is usually employed to lowering costs, boosting efficiency, and enhancing production. Among these processes, production and logistics is of special interest as decisions on supply chains, factory placements, and production allocation are of increasingly more complexity in an interconnected world. Many of these problems can be cast as combinatorial optimization, which are known to be hard to solve as the number of variables increases. Solving these problems at large scale presents a challenge for current computation paradigms, that is, classical super computers.

Quantum computing promises to help dealing with
limits of classical computing. In fact, many algorithms for solving combinatorial problems have been proposed, for instance, the \ac{qaoa} \cite{farhi2014quantum}, recursive QAOA (rQAOA) \cite{Bravyi2020, Bravyi2022hybridquantum}, linear-ramp \ac{qaoa} (LR-QAOA) \cite{montanezbarrera2024universalqaoaprotocolevidence}, adiabatic quantum optimization or quantum annealing \cite{RevModPhys.90.015002}, and counter adiabatic QAOA (CA-QAOA) \cite{Wurtz2022counterdiabaticity}, which could help alleviating the scaling limitations of classical algorithms. For a review of \ac{qaoa} and its variants see \cite{BLEKOS20241} and for a practical description of problem formulations see \cite{10.3389/fphy.2014.00005}. Despite recent progress, it is still unclear whether quantum optimization will be able to solve industry relevant problems. In fact, few classes of problems are known to have guaranteed quantum advantage \cite{montanaro2024quantumspeedupssolvingnearsymmetric} and others show promising scaling \cite{doi:10.1126/sciadv.adm6761}. Most of these algorithms have been implemented in quantum computers for small problem instances with varying levels of success. For example a variant of \ac{qaoa} has been used to solve instances of the Max-Cut problem up to 127 qubits in quantum hardware \cite{sachdeva2024quantumoptimizationusing127qubit}. Currently, due to cloud access to quantum computers, researchers are able to routinely test and deploy optimization algorithms in small scale quantum computers, so-called NISQ (noisy intermediate-scale quantum) \cite{preskill2018, RevModPhys.94.015004} devices. 

\begin{figure}
    \centering
    \includegraphics[width=\linewidth]{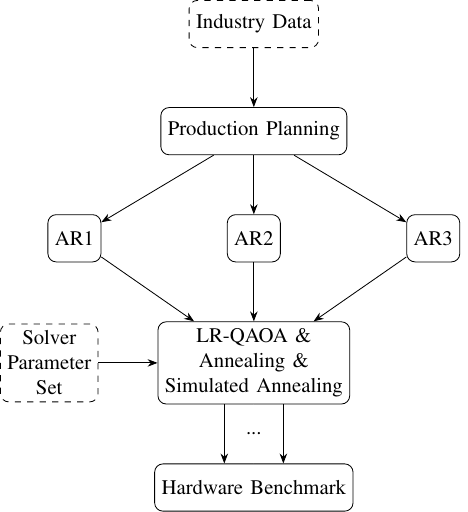}
    \caption{Processing stages overview. Algebraic representations (AR) devised from the Production Planning use case are cast into hardware executable representations for \ac{lrqaoa}, (quantum) annealing and simulated annealing.}
    \label{fig:Intro_Overview}
\end{figure}

Most experiments and demonstrations of quantum algorithms are carried out with a reduced class of problems, for instance, Max-Cut, Max-SAT \cite{10.1145/227683.227684}, or traveling salesperson problem (TSP) \cite{63205816-165a-3cfa-bc71-76de0e788f1a}, which are meant to be representative problems (and suitable abstractions) that are hard to solve classically. 
However, industrial problems seldom fall exactly in a standard category. 
In this work, we compare the performance of quantum and simulated annealing, as well as the \ac{lrqaoa}, ran in quantum hardware, when deploying an industry-relevant production and logistics use case: the capacity planning for production of the body parts of a vehicle. 
Our problem is formulated with real production data and real-world constraints, which in general differ from standard classes of optimization problems. 
\autoref{fig:Intro_Overview} gives an overview of the general data flow in terms of abstraction layers.
While the real industry data forms the input of the parameterized production planning use case, there are a multitude of possibilities to cast that problem into a hardware executable form.
We compare different problem encodings (see \autoref{fig:Intro_Overview}: AR1, AR2, AR3) and investigate how the quality of solutions varies with them.
In particular, we vary the formulation of penalty constraints and data refinement, while also considering different solver specific parameters.
We aim to bridge the gap in benchmarking quantum hardware including non-standard, industry-relevant problems.

The remainder of the paper is divided as follows: While \autoref{sec:related_work} discusses the current state of the art, \autoref{sec:fundamentals} introduces the necessary algorithmic building blocks.
We go into detail about the formulation of the industry relevant use case in \autoref{sec:usecase}.
Although we focus on 3 different algebraic representations, experiments have many more free parameters to optimize for cost, solution quality or time-to-solution. \autoref{sec:experimentalSetup} describes which parameters form the basis for the empirical evaluation in \autoref{sec:experimentalResults}.
\autoref{sec:concl} concludes our study and suggests future improvements.

This paper is augmented by a comprehensive \repro (link in PDF) that also allows for extending our work. 
It also provides additional detailed figures about our experiments.

\section{Related Work}
\label{sec:related_work}
Nenno and Caspari \cite{Nenno_2024} analyze the general steps needed for using quantum computers on dynamic optimization problems with respect to an industry-relevant simplified chemical reactor by incorporating a \ac{qubo} formulation. 
The \ac{qubo} is based on a system of differential-algebraic equations that is embedded into the optimization problem---opposed to former methods that require analytic or parametric solutions \cite{Deng_2023, fernandezvillaverde_2023}.

Formulating problems as \ac{qubo} or more general as pseudo boolean functions (\ie, polynomials) is not a trivial task.
For example, Häner \etal~\cite{haener_2018} outline the challenges for evaluating non-polynomial functions on quantum hardware.
Apart from implementation challenges, highly industry relevant join-ordering problems and their performance-critical formulation into \ac{qubo} form for database query optimization can be found in \cite{schoenberger:23:pvldb}.

To map such problems into \ac{qubo} form, it is necessary to identify and map (in-)equality constraints, discretize continuous variables, apply Lagrange factors, etc.. 
Glover \etal~\cite{Glover_2019} provide an extensive study on transforming optimization problems (\eg, quadratic assignment, knapsack, constraint satisfaction or max-cut) into \ac{qubo} form. 
From the point of view of a real application, mapping to \ac{qubo} requires identifying unambiguous similarities to optimization problems.
For example, Schütz \etal~\cite{Schuetz_2022} show how to cast robot trajectory planning into \ac{qubo} form.
Apart from direct \ac{qubo} mappings, firstly mapping to \ac{pubo} and then transforming to \ac{qubo} can be a valid choice that, however, impacts non-functional requirements and changes problem specific characteristics. 
We shed light on this method for a Job-Shop Scheduling problem by using automatic means of transformation \cite{Schmidbauer_2024}.
These algebraic representations form the basis for further transformations down to a hardware executable representation (see \autoref{fig:Intro_Overview})---for instance, selecting a solver strategy.

Hauke \etal~\cite{Hauke_2020} and Yarkoni \etal~\cite{Yarkoni_2022} give a broad overview of industrial applications  and perspectives using quantum annealing---including traffic flow, scheduling, quantum simulation and finance.
A recent work by Vandelli \etal~\cite{Vandelli2024} simulates \ac{qaoa} and annealing for power consumption in telecommunication networks for up to $31$ qubits.
They show that finding near-optimal solutions is possible---even for constrained problems with $31$ qubits.
Krol \etal~\cite{krol2024qissquantumindustrialshift} present a quantum version of an industrial shift scheduling problem via another solving strategy, that is, Grover's search.

\section{Fundamentals}
\label{sec:fundamentals}
\subsection{LR-QAOA}
\ac{qaoa}, originally proposed by Farhi \etal~\cite{farhi2014quantum}, is a hybrid quantum classical algorithm.
Apart from using \ac{qaoa} as a solver for \acp{op}, Morales \etal~\cite{Morales_2020} consider the conditions for universal computations.
The quantum circuit for \ac{qaoa} consists of $p$ layers of a problem specific part $H_C(\gamma_i)$ and a mixer $H_M(\beta_i)$, parametrized by rotation angles $\gamma_i$ and $\beta_i$. 
These angles are subject to a classical optimizer, which requires multiple executions of the quantum circuit to find a (local) minimum of an optimization problem, encoded in $H_C$.

In contrast to \ac{qaoa}, \ac{lrqaoa} \cite{montanezbarrera_2024} uses predetermined values for rotation angles in each layer. 
The values for $\gamma_i$ (the problem specific part) increase linearly, while the values for $\beta_i$ (the mixer) decrease linearly with each layer, respectively.
However, this necessitates normalizing the objective function, which is computationally feasible in polynomial time.
Since quantum annealers can also realize linear schedules, comparing them to \ac{lrqaoa} is interesting in view of industrial use cases in terms of performance, solution quality, resource requirements, scaling behavior, as well as characterizing the impact of problem formulations on these measures.

\subsection{Adiabatic Quantum Computing}
Contrary to discretized circuit-based models of quantum computation, adiabatic quantum computing is a continuous process. 
Nevertheless, both models of computation are polynomially equivalent in their computational power~\cite{Aharonov_2008}.
When slowly evolving a time-dependent Hamiltonian $H(\tau)$ ($\tau \in [0,T]$), while starting in the ground state at $\tau=0$, the ground state is preserved at $\tau=T$ with probability $P$.
The probability $P$ is close to $1$ when the spectral gap $\delta(\tau)$ (\ie, the absolute eigenvalue difference between the first excited state and the ground state\footnote{We assume a non-degenerate ground state of $H(\tau)$.}) is strictly greater than $0$ ($\forall \tau \in [0,T]$) and the process evolves slowly~\cite{McGeoch_2014}:
Let $n$ denote the problem size and let $\delta_m$ denote the minimum spectral gap.
Then, the time $T$ is polynomial in $n$, iff $\delta_m$ is inverse polynomial in $n$.

The time-dependent Hamiltonian
\begin{equation}
\label{eq:QA_HamiltonianSplit}
    H(\tau) = \frac{T-\tau}{T} H_\text{init} + \frac{\tau}{T} H_\text{final}
\end{equation}
can be split into an easy to prepare initial Hamiltonian $H_\text{init}$ (with known ground state) and a Hamiltonian $H_\text{final}$ that encodes the optimization problem.
While \autoref{eq:QA_HamiltonianSplit} interpolates linearly between $H_\text{init}$ and $H_\text{final}$, other annealing schedules are possible.
Unfortunately, finding the minimum spectral gap to adjust the schedule is a hard problem with recent development in terms of mitigating anti-crossings \cite{Choi_2020, Braida_2021, Fujii_2023}.

The D-Wave Quantum Annealer is a non-universal quantum computer in the sense that not every possible state in the Hilbert Space can be reached.
This is a result of the time-dependent Hamiltonian that is realized in hardware \cite{Zaborniak_2021}:
\begin{align}
    H &= - \frac{A(\tau)}{2} \sum_{i} \sigma_x^{(i)} \label{eqs:DWave_Hamiltonian_Init}\\
      &+ \frac{B(\tau)}{2}\left[ \sum_{i} h_i \sigma_z^{(i)} + \sum_{i < j} J_{ij} \sigma_z^{(i)}\sigma_z^{(j)} \right] \label{eqs:DWave_Hamiltonian_Problem},
\end{align}
where $\sigma_x^{(i)}$ and $\sigma_z^{(i)}$ denote the Pauli-X and -Z operations applied to qubit $i$, respectively.
Similarly to \autoref{eq:QA_HamiltonianSplit}, $A(\tau)$ and $B(\tau)$ adjust the influence of the initial  Hamiltonian (\autoref{eqs:DWave_Hamiltonian_Init}) and the problem Hamiltonian (\autoref{eqs:DWave_Hamiltonian_Problem}).
In particular, we can choose $A(\tau)$ and $B(\tau)$ to represent an approximately linear annealing schedule.
Although the D-Wave Quantum Annealer cannot reach any possible state in Hilbert Space, it is a universal model of computation in the sense of solving NP-complete problems due to solving \acp{qubo}. 
As a side note, by exploiting level crossings, one can implement gate operations beyond $\sigma_z$ and $\sigma_x$ \cite{imoto_2024}.

\subsection{Simulated Annealing}
Simulated annealing is a probabilistic classical method to solve optimization problems \cite{Nikolaev_2010} and in particular \acp{pbf}.
Therefore, this method can serve as a classical baseline for experiments.
In essence, one chooses a random initial state and then repeats the following $n$ times:
Firstly, choose a random variable in the current bit string and flip it\footnote{Deterministically choosing the next variable is also possible.}. 
If this flip lowers the energy of the \ac{pbf}\footnote{For minimization problems.} $f$, accept that flip.
Conversely, if this flip increases the energy of $f$, accept this flip based on a random experiment.
The probability to accept the flip depends on the energy difference as well as a typically decreasing base probability \cite{McGeoch_2014}.
For example, one can choose the probability $P$ as $P = \min(1, e^{-(\Delta/i)})$,
where $\Delta$ denotes the energy difference to the function without the bit flip and $i$ is the iteration count.
If the random experiment does not accept the flip, revert it.

\section{Use Case Modelling}
\label{sec:usecase}
\begin{figure}
    \centering
    \includegraphics[width=\linewidth]{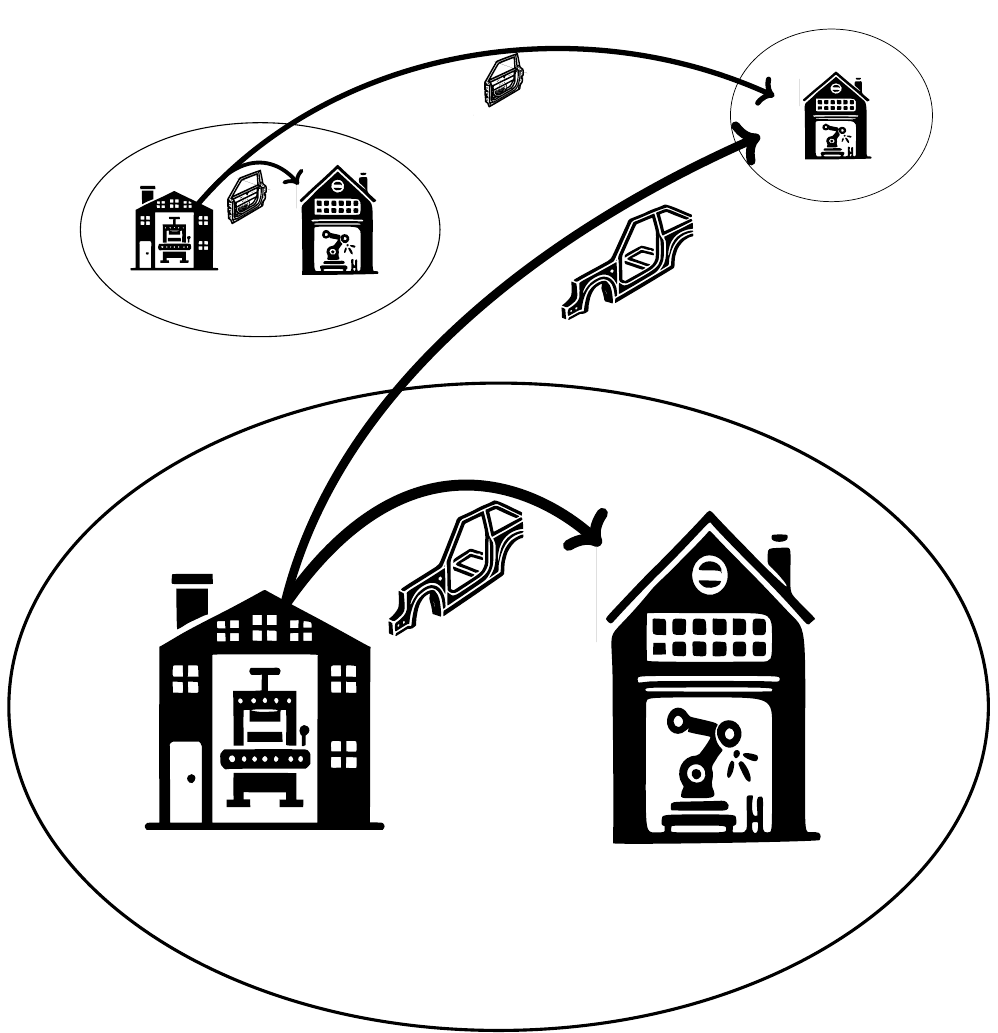}
    \caption{An exemplary scheme of three production sites, two comprising a press shop and an assembly plant, one with an assembly plant only, sized to convey an impression of distance (smaller means farther away). Toolkits are assigned to the press machines of the press shops, thus determining where which parts are pressed and where consequently the flows of produced parts to the assembly plants must originate from. Product flows of a door and a side-frame can be seen, each fulfilling onsite as well as offsite demands.}
    \label{fig:sites}
\end{figure}

\subsection{Description and Mathematical Formulation}

In the complex process of vehicle mass production, forming of raw materials into product parts is an early and fundamental step.
One of such processes is done in the \textit{press shop}, in which metal sheets are shaped into the components of the body of a vehicle. Press shops may house several \emph{press machines}, which provide the force by which the metal is reshaped and cut using \emph{toolkits}, which are the tooling components with the shape of body parts. Press shops are distributed globally and do not necessarily coincide with the assembly plants, which put together the whole vehicle. Production and logistic costs are associated with every choice of production allocation as well as legal and technical constraints regarding volume of parts which can be produced at a given time and location. For example, each press machine has a maximum capacity it can offer during a given production period and each toolkit must produce a given number of parts. Figure \ref{fig:sites} gives an illustration of the setting.

The question for long-term planning is to decide which toolkits should be assigned to which press machines in a way that is cost optimal. The decisions, for example, how many parts to produce or which toolkit to use for the production of which part, are given parameters and not degrees of freedom in finding optimal solutions. Therefore, each toolkit can be assumed to ``carry'' a given amount of parts demanded, which, via press-machine dependence and given work rates, translates into an effective (possibly press-machine dependent) workload. Thus, the allocation decision can be cast as an optimization problem in which the total production costs are minimized. 
The minimization is subject to all parts being produced while respecting the maximum capacities of every machine.

This optimization problem can be formulated as a binary constrained linear program. We define $T$ as the set of all toolkits and $M$ as the set of all machines. Then, we can introduce the decision variable $x_{tm}$ to encode which toolkit $t \in T$ is assigned to which machine $m \in M$,
\begin{equation}\label{eq:decision_variable}
    x_{tm} = \begin{cases}
        1 & \text{if toolkit $t$ is assigned to machine $m$,} \\
        0 & \text{otherwise.}
    \end{cases}
\end{equation}
The cost arising by an assignment is given by $c_{tm}$, leading to the objective function
\begin{equation}\label{eq:objective}
    \min_{\mathbf{x}} \sum_{t \in T} \sum_{m \in M} c_{tm} \; x_{tm},
\end{equation}
where $\mathbf{x} = (x_{tm})_{t \in T, m \in M} \in \{0,1\}^{T \times M}$.
Each machine has a maximum capacity $h_m$, usually given in units of hours per production period, and each toolkit assignment requires a demanded workload $w_{tm}$, also in hours per production period, resulting in the capacity constraints
\begin{equation}\label{eq:capacity_constraint}
    \sum_{t \in T} w_{tm} \; x_{tm} \leq h_m, \phantom{space} \forall \; m \in M.
\end{equation}
Moreover, each toolkit has to be assigned to a machine exactly once, which is encoded as
\begin{equation}\label{eq:exactly_once_constraint}
    \sum_{m \in M} x_{tm} = 1, \phantom{space} \forall \ t \in T. 
\end{equation}
This constraint ensures that each toolkit is assigned to exactly one location and also prevents that nothing is produced despite of existing demands. Note that producing nothing would be the cheapest option concerning the objective \autoref{eq:objective}.

We use this linear program both for deriving the problem formulation for the quantum computer and for computing the optimal solutions of the problem instances in Sec.~\ref{sec:experimentalSetup}. For the latter, we use the SCIP -- short for \textit{S}olving \textit{C}onstraint \textit{I}nteger \textit{P}rograms -- solver \cite{SCIP} via the pywraplp interface from Google OR-Tools \cite{ortools}. This is done for comparison purposes.

\subsection{QUBO Creation and Penalty Terms}
The standard way to run an optimization problem in a quantum computer is to reformulate it as a \ac{qubo}. The procedure of transforming the original, constrained, optimization (\autoref{eq:objective}, \ref{eq:capacity_constraint} and \ref{eq:exactly_once_constraint}) into an unconstrained problem is to add the constraint violations as penalty terms to the objective function (\autoref{eq:objective}). For our reformulation, we use the Qiskit optimization library (version 0.6.0) \cite{qiskit2024}. In the following, we sketch the resulting steps.

First, inequality constraints, in our case the capacity-constraints (\autoref{eq:capacity_constraint}), are transformed into equality constraints. The transformation is achieved by introducing so-called ``slack variables'', which we denote by $S_m$ since there is one capacity constraint for each press machine. The reformulated capacity constraints read
\begin{equation}\label{eq:rescaled_capacity_constraint}
    \sum_{t \in T} w_{tm} \; x_{tm} + S_m - h_m = 0, \phantom{space} \forall m \; \in M,
\end{equation}
where $0 \leq S_m \leq h_m$, since the lower bound of $\sum_{t \in T} w_{tm} \; x_{tm}$ is zero. Next, the newly introduced slack variables $S_m$ have to be split into a binary representation to fulfill the requirements of a \ac{qubo}, 
\begin{equation}\label{eq:binary_slack}
    S_m = \sum_{j=0}^{r_m-1} 2^j s_{mj} + (h_m - 2^{r_m} + 1) s_{m r_m}, 
\end{equation}
extending the problem by the binary variables $s_{m0}, \dots, s_{m r_m} $ per press machine $m$, where $r_m = \lfloor\log_2(h_m)\rfloor$.

Now, all constraints are equalities and can be transformed into penalties of parabolic shape that are added to the objective function. For the assignment constraints (\autoref{eq:exactly_once_constraint}) we obtain the penalty terms
\begin{equation}
    \lambda_t \left( \sum_{m \in M} x_{tm} - 1 \right)^2, \phantom{space} \forall \ t \in T,
\end{equation}
with penalty factors $\lambda_t$ and for the capacity constraints (\autoref{eq:capacity_constraint}) we write the penalty terms
\begin{equation}
    \lambda_m \left( \sum_{t \in T} w_{tm} \; x_{tm} + S_m - h_m \right)^2, \phantom{space} \forall m \; \in M,
\end{equation}
with penalty factors $\lambda_m$ and $S_m$ substituted by \autoref{eq:binary_slack}. The total \ac{qubo} problem is then given by the objective function
\begin{equation}\label{eq:qubo_objective}
    \begin{split}
        \min_{\mathbf{x}, (\mathbf{s}_m)_{m \in M}} &\left[ \sum_{t \in T} \sum_{m \in M} c_{tm} \; x_{tm} \right. + \sum_{t \in T} \lambda_t \left( \sum_{m \in M} x_{tm} - 1 \right)^2 \\\
        & \phantom{+} \left. + \sum_{m \in M} \lambda_m \left( \sum_{t \in T} w_{tm} \; x_{tm} + S_m - h_m \right)^2 \right],
    \end{split}
\end{equation}
with $S_m$ given by the binary variables $s_{mj}$ defined in \autoref{eq:binary_slack}, combined to the binary vector $\mathbf{s}_m = (s_{m0}, \dots, s_{m r_m})$. The \ac{qubo} problem (\autoref{eq:qubo_objective}) can also be brought into matrix form,
\begin{equation}
\label{eq:qubo_minimization}
    \min_{\mathbf{x} \in \{0,1\}^N} \mathbf{x}^\mathrm{T} \mathbf{\hat{Q}} \mathbf{x} + C,
\end{equation}
where the binary vector $\mathbf{x}$ collects all decision variables $x_{tm}$ and slack variables $s_{mj}$, and the constant $C$ subsumes all constants. Since $x^2 = x$ for binary variables $x$, all coefficients of the terms of degree 1 and 2 in the variables $x_{tm}$ and $s_{mj}$ can be collected in the \ac{qubo} matrix $\mathbf{\hat{Q}}$. 

So far, we have not discussed how to choose the penalty factors $\lambda_t$ and $\lambda_m$. Moreover, there are different possibilities for encoding the data of actual problem instances into the linear program and thus the \ac{qubo}. In Sec.~\ref{sec:experimentalSetup} we present three different approaches for building the \ac{qubo} which we benchmark on quantum hardware in Sec.~\ref{sec:experimentalResults}. 

\section{Experimental Setup}
\label{sec:experimentalSetup}

\subsection{Problem instances}
For our experiments, we create 6 problem instances from anonymized
production data. These problem instances are small enough to fit in a quantum computer, but the values of costs and capacities are taken to represent real situations. 
That means that the costs, capacities, and restrictions are quantitatively comparable with real production data. The only adjustment we make is to safely round the values entering the capacity constraint to integers, that is, the maximum machine capacity $h_m$ is rounded down and the capacity $w_{tm}$ required by a toolkit assignment is rounded up. We summarize the problem instances in \autoref{tab:problem_instances}.

\begin{table}[h!]
    \centering
    \caption{Problem instances.}\label{tab:problem_instances}
    \begin{tabular}{ccc}
        \toprule
        \textbf{Toolkits} & \textbf{Machines} & \textbf{Qubits} \\
        \midrule
         3 & 2 & 22 \\
        \rowcolor{gray!10} 9 & 2 & 36 \\
        13 & 2 & 46 \\
        \rowcolor{gray!10} 16 & 2 & 54 \\
        18 & 2 & 58 \\
        \rowcolor{gray!10} 19 & 2 & 60 \\
        \bottomrule
    \end{tabular}
\end{table}

All problem instances are represented by 3 different \ac{qubo} encodings (see \autoref{fig:Intro_Overview}: AR1, AR2, AR3):
\subsubsection{Raw QUBOs}
We take the data directly without any processing and create \ac{qubo} formulations for each problem instance. We do not optimize the penalty factors. Instead, we choose them in a broad grid $\lambda_m\in\{10^{3}, 10^{4}, 10^{5}\}$ and $\lambda_t\in\{10^{7}, 10^{8}, 10^{9}\}$, taking the same values for all machines $m$ and toolkits $t$, respectively.

\subsubsection{Scaled QUBOs}
We adjust the scaling of the assignment constraint by multiplying \autoref{eq:exactly_once_constraint} with $\lambda_s \in \{0.1, 1\}$. Moreover, before transforming the constraints to penalties but after transforming inequalities to equalities, we rescale the constraints from \autoref{eq:exactly_once_constraint} and \autoref{eq:rescaled_capacity_constraint} such that they obtain a similar value range as the objective function in \autoref{eq:objective}. We define the value range $v$ of a term $f$ as 
\begin{equation}\label{eq:value_range}
    v = \left| \lceil f \rceil - \lfloor f \rfloor \right|.
\end{equation}
For the rescaling, we first determine the largest value range $v_{\max}$ occurring in the linear program by evaluating \autoref{eq:value_range} for the objective function in \autoref{eq:objective} and the left-hand side of the constraints in \autoref{eq:rescaled_capacity_constraint} and \autoref{eq:exactly_once_constraint} (after multiplying \autoref{eq:exactly_once_constraint} with $\lambda_s$). Then, we divide both the objective function, \autoref{eq:objective}, and all constraints, \autoref{eq:exactly_once_constraint} and \autoref{eq:rescaled_capacity_constraint}, by their respective range $v$ and then multiply them by $v_{\max}$. Afterwards, we proceed with the transformation of the linear program to a \ac{qubo} by transforming the constraints to penalties. The rescaling of the constraints replaces optimizing the penalty factors, such that $\lambda_m = \lambda_t = 1$ for all $m \in M$ and $t \in T$.

\subsubsection{Rounded-cost QUBOs}
For this formulation, we first take the data and rescale the production costs so that the minimum cost is 1. To avoid non-integer cost values, we employ integer division for this rescaling. Then, we proceed with the same transformation routine as described for the \textit{scaled} \acp{qubo}. The aim of rescaling the costs is that the resulting \ac{qubo} matrix should be more balanced than for the first two formulations---meaning that the difference between the minimum and maximum value in the \ac{qubo} matrix is significantly reduced.

\subsection{Methods}
\label{ssec:Methods}
Evaluating multiple problem instances with multiple solver strategies and (quantum) hardware represents paths in the abstraction layer graph (compare to \autoref{fig:Intro_Overview}). 
On top of choosing a path, free parameters in transformations influence properties and ultimately performance. 
We always test all mentioned problem sizes (see \autoref{tab:problem_instances})---independently of the used solver strategy and hardware.
Additionally, depending on the used problem instance (\textit{raw}, \textit{scaled} and \textit{rounded}), we test their respective penalty factors for all solver strategies.

For the solver strategy \ac{lrqaoa}, we identify the number of layers $p$ and the variation in the slope defining parameters $\Delta_{\gamma}$, $\Delta_{\beta}$ (see \autoref{sec:fundamentals}) as important parameters.
Preliminary tests on hardware indicated that any $p > 10$ does not lead to improved performance. 
Hence, we select $p \in \{1,2,5,10\}$.
Montanez-Barrera and Michielsen \cite{montanezbarrera_2024} test $\Delta_{\gamma}$, $\Delta_{\beta}$-dependent performance for six optimization problems. 
Based on their results for pure optimization problems, we choose $\Delta_{\gamma} = 0.9$ and $ \Delta_{\beta} = 0.6$.
Any logical quantum circuit can be depth-optimized in polynomial time---only differing from the optimal solution by at most one \cite{Broadbent_2009}.
The edge coloring problem and Vizing's theorem \cite{diestel_2005} provide the basis for this argument.
However, optimizing circuit depth becomes NP-hard, when considering restricted arbitrary hardware topologies.
For specific topologies there have been advances~\cite{kattemolle_2024}, in particular for the heavy-hex topology~\cite{Kim_2023}.
We employ an almost depth optimal logical circuit (via edge coloring), which is then transpiled to the hardware gate set and topology of \texttt{ibm\_marrakesh}. 
Qiskit~\cite{qiskit_2024} provides four levels of optimization of which we use level $0$ (no optimization) and $3$ (highest optimization).

For the solver strategy of quantum annealing, recall that annealing-based approaches are influenced by the minimum spectral gap (see \autoref{sec:fundamentals}).
Hence, the annealing time is an important parameter to avoid level crossings. 
Therefore, we choose the annealing time $\tau \in \{10, 20, 40, 80, 160, 320, 640, 1280\}[\mu s]$ for the \texttt{Advantage\_system4.1} QPU that covers $5760$ qubits via the Pegasus topology \cite{dattani2019pegasussecondconnectivitygraph}.
Note that the maximum allowed annealing time also depends on the number of shots, which we choose to be $500$ to improve measurement statistics.
Additionally, a custom annealing schedule can accelerate annealing where the spectral gap is large and decelerate annealing where the spectral gap is small. 
Since solving for an optimal annealing schedule is at least as hard as solving the optimization problem at hand\footnote{We would need to know the spectral gap for every $\tau$.}, we use $4$ annealing schedules:
\begin{enumerate}
    \item \textit{Linear}: \resizebox{\baselineskip*3/5}{!}{\tikz\draw (0,1) -- (1,0);}
    \item \textit{Bowover}: \resizebox{\baselineskip*3/5}{!}{\tikz\draw (0,1) .. controls (0.4,1) and (1,0.4) .. (1,0);}
    \item \textit{Bowunder}: \resizebox{\baselineskip*3/5}{!}{\tikz\draw (0,1) .. controls (0,0.6) and (0.6,0) .. (1,0);}
    \item \textit{SteepFlatSteep}: \resizebox{\baselineskip*3/5}{!}{\tikz\draw (0,1) .. controls (0,0.55) and (1,0.45) .. (1,0);}
\end{enumerate}
Picture a linearly decreasing schedule $s_\textit{Linear}(\tau)$ \footnote{A decreasing schedule refers to $H_\text{init}$. $H_\text{final}$ is scaled accordingly.}.
Then, \textit{Bowover} represents a schedule above $s_\textit{Linear}(\tau)$---leading to slow annealing at the start and faster annealing when progressing.
Analogously, \textit{Bowunder} represents a schedule below $s_\textit{Linear}(\tau)$---having the exact opposite effect.
Finally, \textit{SteepFlatSteep} is a combination of \textit{Bowunder} at the start and \textit{Bowover} at the end---leading to slower annealing in the middle of the process.
To map a given problem instance onto D-Wave's hardware, we use the heuristic MinorMinor embedding \cite{cai_2014}:
It finds a minor-embedding of the graph representation of a given \ac{qubo} instance in the graph that represents D-Wave's hardware (see \cite{Dvorak2025} for more information on graph minors).
Note that an embedding for a \ac{qubo} of size $|Q|$ usually requires more than $|Q|$ qubits in D-Wave's hardware.

Similar to the time parameter in quantum annealing, simulated annealing can employ longer classical runtime (\ie, steps $n$; see \autoref{sec:fundamentals}) to heuristically refine the current local minimum.
Also, the probability function can be changed to, for instance, overcome local minima.
However, we use simulated annealing as a baseline and thus use the standard geometric schedule and set $n = 1280$.

To accommodate for a single bit-flip error, we use a classical post-processing error mitigation technique for all experiments (see \cite{montanezbarrera_2024}): For each bit $x$ in a sampled bitstring $\mathbf{x}$, we test if the negation of $x$ improves the energy with respect to the \ac{qubo} (see \autoref{sec:experimentalSetup}). 
We then use $\mathbf{x}'$ with lowest energy. 
Note that its runtime scales linearly in the size of the bitstrings and the number of samples.

\subsection{Expected Behavior}

It is evident that current \ac{nisq}-era hardware still is restricted by noise. 
Although \acp{qecc} are in development and have provable advantages, their use is to the detriment of requiring more qubits (see \cite{Ball_2023}). 
\acp{qecc} offer potential to allow for some degree of noise, while protecting the intended quantum state.
Hence, (depending on the number of additionally used qubits), \acp{qecc} will eventually make noisy quantum hardware behave similar to noiseless systems and noiseless simulations.
For the following experimental analysis, we expect bigger problem sizes to perform worse, due to either exponentially scaling search spaces or noise that arises from inherently longer circuits (\ac{lrqaoa}) and bigger embedding size (annealing).

As the number of layers $p$ in \ac{lrqaoa} increases, the logical circuit depth increases linearly with $p$. 
While noiseless \ac{lrqaoa} converges to an optimal solution (under the assumptions of the adiabatic theorem; see \autoref{sec:fundamentals}), hardware noise eventually leads to a (non-uniform\footnote{This is hardware dependent (\eg, via topology).}) random output distribution. 
We confirm the convergence to optimal solutions numerically by simulating the $3$ Toolkit ($22$ qubit) case for up to $p=100$ \ac{lrqaoa} layers. 
Notably, even for $p=1$ and $1000$ shots, we find the optimal solution regardless of the problem variant (\textit{raw}, \textit{scaled} or \textit{rounded}).
However, we notice that the \textit{rounded} variant performs better for $p \geq 5$ than the \textit{raw} and \textit{scaled} variant.
The \repro (link in PDF) contains detailed figures.
Take into consideration that system noise eventually diminishes better Trotterization accuracy (\ie, higher $p$). 
Noise can also blur variant-specific effects.
Therefore, the free parameter $p$ is subject to balancing both noise and approximating annealing through Trotterization.
Hence, a (variant-dependent) tipping point should be evident, when increasing $p$. 
A similar effect can be expected for the annealing time and the annealing schedule to avoid level-crossings in the adiabatic evolution. 
They are also free parameters that need to balance the expected noiseless performance gain and the effects of noise. 

Concerning the different \ac{qubo} formulations, \textit{raw}, \textit{rounded}, and \textit{scaled}, the number of effective hardware runs differs for each of the variants due to the \ac{qubo}'s respective penalization strategy. For \textit{raw} we have $9$ combinations; for \textit{scaled} $2$ combinations, and for \textit{rounded} $1$ combination.
If our experiments were mere random sampling and all effects of the quantum circuit on the final state measurement were washed out by noise, we would expect the \textit{raw} \ac{qubo} to perform best simply due to being run more often than the other versions and thus having the highest probability of sampling a good result eventually. 
Of course, we expect the quantum circuit to have a measurable effect on the final quantum state in our experiments and the different penalization strategies alter the \ac{qubo} formulation and should hence perform differently. Nevertheless, the different number of runs could result in a (slight) bias in favor of the \textit{raw} variant.

\section{Experimental Results}
\label{sec:experimentalResults}
For each of the following combined figures, we show results for quantum annealing on top and results for \ac{lrqaoa} below.
Each figure has the problem size (\ie, the number of toolkits) on its (non-equally-spaced) x-axis.
Note that toolkits directly translate to number of qubits, since we always use 2 press machines (see \autoref{tab:problem_instances}).
Data points in each figure must adhere to its respective criteria (\eg, being valid). 
Consequently, if there are no data points for toolkits $t$, we omit $t$ in the figure.

\begin{figure}[htb]
    \centering
    \includegraphics[width=\linewidth]{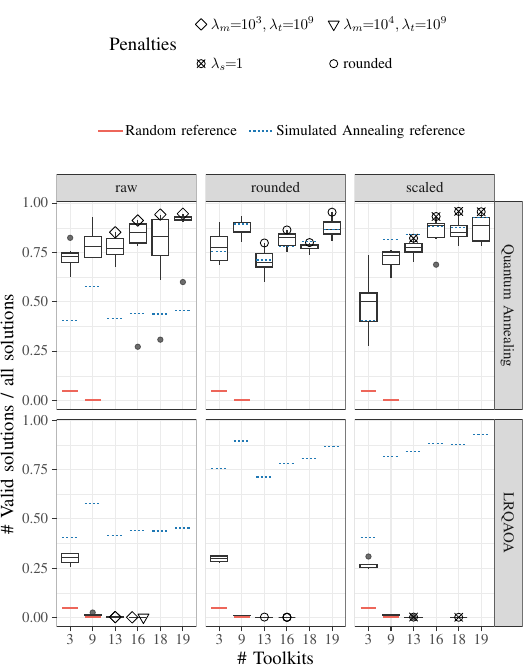}
    \caption{Problem size (x-axis) vs percentage of valid (\ie, constraint satisfying; see \autoref{sec:usecase}) solutions (y-axis) and corresponding best performing penalties (see \autoref{sec:experimentalSetup}). Closer to $1.00$ is better. Horizontal facets: quantum annealing on D-Wave and \ac{lrqaoa} on IBM. Vertical facets: problem formulation (see \autoref{sec:usecase} and \autoref{fig:Intro_Overview}). \textcolor{lfdred}{Random} (solid line) and \textcolor{lfdblue}{Simulated Annealing} (dotted line) as reference (same values for both horizontal facets).}
    \label{fig:PercentValid}
\end{figure}

\begin{figure}[htb]
    \centering
    \includegraphics[width=\linewidth]{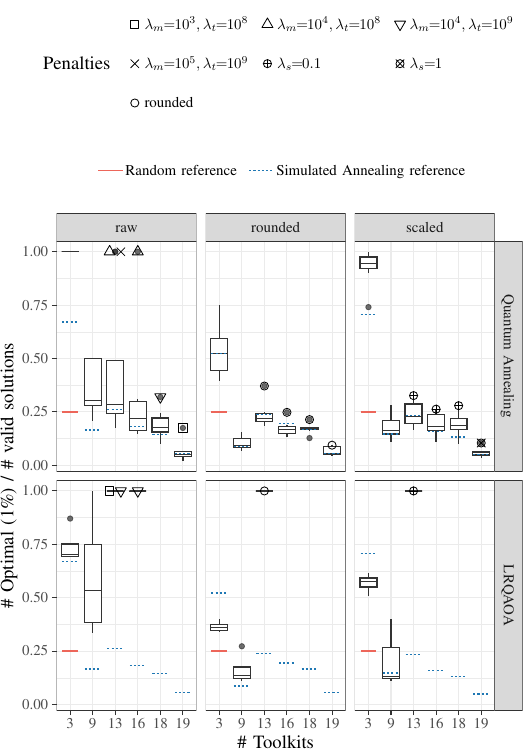}
    \caption{Problem size (x-axis) vs ratio of valid (\ie, constraint satisfying; see \autoref{sec:usecase}) solutions within $1$\% of the optimum to valid solutions (y-axis) and corresponding best performing penalties (see \autoref{sec:experimentalSetup}). Closer to $1.00$ is better. Horizontal facets: quantum annealing on D-Wave and \ac{lrqaoa} on IBM. Vertical facets: problem formulation (see \autoref{sec:usecase} and \autoref{fig:Intro_Overview}). \textcolor{lfdred}{Random} (solid line) and \textcolor{lfdblue}{Simulated Annealing} (dotted line) as reference (same values for both horizontal facets).}
    \label{fig:PercentOptValid}
\end{figure}

\begin{figure}[htb]
    \centering
    \includegraphics[width=\linewidth]{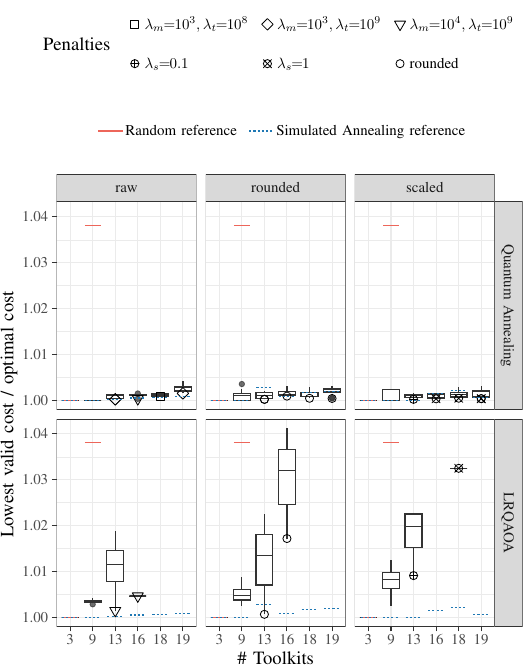}
    \caption{Problem size (x-axis) vs normalized lowest cost for valid (\ie, constraint satisfying; see \autoref{sec:usecase}) solutions (y-axis) and corresponding best performing penalties (see \autoref{sec:experimentalSetup}). Closer to $1.00$ is better. Horizontal facets: quantum annealing on D-Wave and \ac{lrqaoa} on IBM. Vertical facets: problem formulation (see \autoref{sec:usecase} and \autoref{fig:Intro_Overview}). \textcolor{lfdred}{Random} (solid line) and \textcolor{lfdblue}{Simulated Annealing} (dotted line) as reference (same values for both horizontal facets).}
    \label{fig:LowestByOptCost}
\end{figure}

For \autoref{fig:PercentValid}, \ref{fig:PercentOptValid} and \ref{fig:LowestByOptCost}, we pick the best performing penalty for each annealing time or layer $p$ (see \autoref{ssec:Methods}) per toolkit and problem formulation and then show their values in a boxplot.
Its box is bounded by the first and third quartile and additionally shows the median as a line.
Furthermore, its whiskers extend to at maximum $1.5 \cdot \mathrm{IQR}$ (depending on actually available data points), where the interquartile range ($\mathrm{IQR}$) is the distance between the first and third quartile. 
Any data point outside this range is shown as a (partially translucent) circle (\resizebox{\baselineskip*2/5}{!}{\tikz\filldraw [opacity=0.5] (0,0) circle (3pt);}). 
Moreover, we show the best performing penalty for each x-value (\ie, toolkits) explicitly as a shape:
For the \textit{raw} variant, a (rotated) rectangle (\resizebox{\baselineskip*2/5}{!}{\tikz\draw [] (0,0) rectangle (1,1);}, \resizebox{\baselineskip*1/2}{!}{\tikz\draw [rotate=45] (0,0) rectangle (1,1);}) corresponds to $\lambda_m = 3$ and a rotated triangle (\resizebox{\baselineskip*1/2}{!}{\tikz\node[isosceles triangle,
	isosceles triangle apex angle=60,
	draw,
	rotate=210] (T2) at (0,0){};},\resizebox{\baselineskip*1/2}{!}{\tikz\node[isosceles triangle,
	isosceles triangle apex angle=60,
	draw,
	rotate=270] (T2) at (0,0){};}) corresponds to $\lambda_m=4$. 
Their respective rotation encodes $\lambda_t$.
The \textit{scaled} variant only has two penalties, which we show as a circle with a cross ($\lambda_s=0.1$) or a circle with an x ($\lambda_s = 1$).
Analogously, we show the \textit{rounded} variant as a circle.
Small problem sizes lead to many best performing penalty weights and hence we omit explicit shapes for $3$ and $9$ toolkits. 
For a comparison to classical methods, we use simulated annealing ($1280$ steps) and the best out of $1000$ randomly generated bitstrings as reference.
Although we test four annealing schedules, as described in \autoref{sec:experimentalSetup}, we only show the linear schedule due to its similarity to \ac{lrqaoa}.
Note that the type of our test schedules only has minor influence on the performance.
Similarly, for \ac{lrqaoa} we test optimization level $0$ and $3$ in Qiskit, but restrict \autoref{fig:PercentValid}, \ref{fig:PercentOptValid}, and \ref{fig:LowestByOptCost} to level $3$.
Take into consideration that optimization level $0$ usually worsens the results---with outliers that probably originate from system noise.

\autoref{fig:PercentValid} shows the percentage of valid solutions (closer to $1.00$ is better). Valid solutions satisfy the problem constraints given in \autoref{sec:usecase} (\ie, capacity constraint, \autoref{eq:capacity_constraint}, and assignment constraint, \autoref{eq:exactly_once_constraint}).
For both \ac{lrqaoa} and quantum annealing and each QUBO variant (\textit{raw}, \textit{rounded}, and \textit{scaled}), we see that a specific set of penalties leads to the highest percentage of valid solutions for bigger problem instances.
Note that we test $9$ penalty combinations for the \textit{raw}, two for the \textit{scaled}, and one for the \textit{rounded} variant, due to their different construction.
\ac{lrqaoa} rarely finds valid solutions for bigger problem instances.
Contrary, annealing performs exceptionally well---finding $\geq 90\%$ valid solutions for bigger problem instances.
Interestingly, the simulated annealing reference is Pearson-correlated \cite{Niven_2012} to the mean of the quantum annealing results: $r_\textit{raw} \approx 0.2736$, $r_\textit{rounded} \approx 0.9297$ and $r_\textit{scaled} \approx 0.9659$.
Hence, the \textit{rounded} and \textit{scaled} variant, seem to have beneficial numerical nature for simulated and quantum annealing, which allows for some degree of extrapolation.

Although \autoref{fig:PercentValid} reduces the set of solutions to valid solutions, they can be numerically far from optimal in terms of production cost.
Hence, \autoref{fig:PercentOptValid} restricts the set of valid solutions to solutions that lie within $1\%$ of the optimal solution (closer to $1.00$ is better). 
A higher share of (near) optimal solutions also leads to faster time-to-solution, which is cost beneficial for computation.
Smaller problem instances have a higher share of optimal solutions---potentially due to the exponentially scaling solution space, see random reference (red) that only finds valid solutions for 3 and 9 toolkits and only close-to-optimal solutions for 3 toolkits. 
\ac{lrqaoa} fails to find (near) optimal solutions for $18$ and $19$ toolkits, while quantum annealing finds (near) optimal solutions for all tested toolkits---although their share decreases with increasing problem size.
As before, we calculate the Pearson correlation \cite{Niven_2012} between simulated annealing and the average quantum annealing results:
$r_\textit{raw} \approx 0.9874$, $r_\textit{rounded} \approx 0.9965$ and $r_\textit{scaled} \approx 0.9960$.
Note that \autoref{fig:PercentOptValid} does not show absolute values and depicts a subset of \autoref{fig:PercentValid}. 
By multiplying the percentage of valid solutions (\autoref{fig:PercentValid}) by the share of (near) optimal to valid solutions (\autoref{fig:PercentOptValid}), one can obtain the percentage of (near) optimal solutions, which further highlights the difference between quantum annealing on D-Wave and \ac{lrqaoa} on current IBM devices.

\autoref{fig:PercentValid} and \ref{fig:PercentOptValid} give insights in the distribution of solutions and therefore allow to extrapolate the time-to-solution or runs required to obtain a sufficiently good solution.
Ultimately, the solution with the least cost is of interest for production.
Therefore, \autoref{fig:LowestByOptCost} gives the ratio for the lowest valid to the optimal cost (closer to $1.00$ is better).
Although the results for quantum annealing all lie within $0.41\%$ of the optimal solution, the difference in problem formulation is apparent. 
At $19$ toolkits, the \textit{raw} variant performs worse than the \textit{rounded} an \textit{scaled} variant. 
For completeness, we calculate the Person correlation \cite{Niven_2012} between the mean value for annealing and simulated annealing \cite{Niven_2012}: $r_\textit{raw} \approx 0.9516$, $r_\textit{rounded} \approx 0.4224$ and $r_\textit{scaled} \approx 0.4184$.
As \ac{lrqaoa} produces little to none valid solutions for increasing problem sizes (see \autoref{fig:PercentValid}), they are increasingly farther from the optimal solution. 
Nevertheless, the final measured state obtained from \ac{lrqaoa} still contains problem specific information, since \ac{lrqaoa} is dramatically better than the random reference, which cannot find valid solutions for more than $9$ toolkits.
Hence, system noise does not render the final state useless.
Upcoming implementations of error correction should therefore make a significant contribution to improving the results.
Although, a higher annealing time has a slight net positive effect on the \ac{qubo} energy, with outliers potentially originating from system noise, we do not find a strong influence of annealing time on \ac{qubo} energy.
Hence, the spectral gap does not appear to be the limiting factor for the tested problem sizes.
This also applies to \ac{lrqaoa} analogously.

\begin{figure}[htb]
    \centering
    \includegraphics[width=\linewidth]{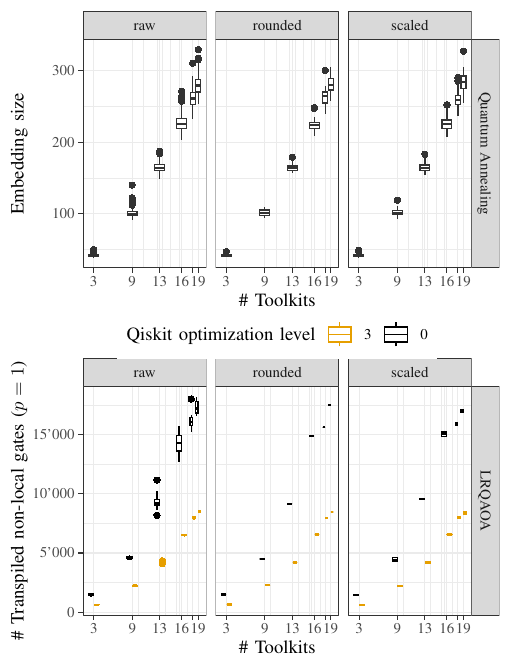}
    \caption{Problem size (x-axis; linearly spaced by \# qubits: see \autoref{tab:problem_instances}) vs scaling behavior (\ie, embedding size or number of transpiled non-local gates for $p=1$; y-axis). Lower is better. Horizontal facets: quantum annealing on D-Wave and \ac{lrqaoa} on IBM. Vertical facets: problem formulation (see \autoref{sec:usecase} and \autoref{fig:Intro_Overview}). Qiskit optimzation level only applies to \ac{lrqaoa}.}
    \label{fig:EmbeddingScaling}
\end{figure}

Overall, quantum annealing outperforms \ac{lrqaoa} for the tested problem instances. 
However, scaling behavior beyond the capabilities of current hardware can put that result into a different perspective.
For \ac{lrqaoa}, we identify the number of transpiled non-local gates (\ie, two qubit gates in our case) as the relevant metric. 
Note that the circuit depth, the number of total gates and the number of used qubits can also be valid metrics.
When considering the effect of transpilation, missing connections in hardware are resolved by introducing additional (non-local) gates. 
Hence, a gate-based approach tends towards deeper circuits, while the number of qubits is not a hard limiting factor: 
Given a quantum hardware with $q_h$ many qubits that form a connected graph.
Then, for any given logical quantum circuit with $q_l \leq q_h$ qubits, there exists a unitary-equivalent (deep) hardware executable circuit (see \cite{Shende_2006} or \cite{Nielsen_2012}).
Contrary, for quantum annealing, we use the embedding size on hardware as the eventually limiting factor, when increasing problem size, since missing connections are resolved by combining hardware qubits \cite{Zbinden_2020}.
\autoref{fig:EmbeddingScaling} shows these metrics on the y-axis. 
The x-axis shows the number of toolkits (we omit the label for $18$ toolkits, but show its data).
It is linearly spaced by the number of qubits, since they correspond to toolkits (see \autoref{tab:problem_instances}). 
In \ac{lrqaoa}, the number of layers $p$ increases the number of non-local gates in the logical circuit linearly.
We found that this is similar for the transpiled circuit---meaning that the transpiler has no significant impact in combining layers. 
Hence, we only show $p=1$ for optimization levels $0$ and $3$.
Optimization level $3$ roughly halves the number of transpiled gates and therefore has significant impact on solution quality that is highly influenced by gate noise.
Compared to \ac{lrqaoa}, quantum annealing has a higher spread in embedding size, with the least spread for the \textit{rounded} variant---albeit there being only minor differences between problem formulations.
Take into consideration that the x-axis is spaced in terms of qubits, but labeled with toolkits (see \autoref{tab:problem_instances}), which allows for comparing the scaling behavior with other industry relevant problems.
Also take into consideration, that the QPU access time is significantly higher with IBM devices and \ac{lrqaoa}, which is relevant for a cost-to-solution estimation.
On average, for a single experiment (500 shots), the D-Wave annealer uses $\approx 0.3$ seconds, while IBM (1000 shots) uses $\approx 7.8$ seconds. 

In summary, the experiments suggest that a choice of penalty weights that clearly outperforms the others emerges with increasing problem size.
Hence, for similarly structured problems, it can be beneficial to first find a set of suitable penalty weights by analyzing smaller but growing problem instances. These optimized weights should then also perform well for larger problem instances.
Moreover, we find similar or better performance for the \textit{rounded} and \textit{scaled} variant over the \textit{raw} variant. 
From an industry point of view, it is therefore not necessary to perform a grid search over the basic formulation, but rather automatically select suitable parameters (as in the \textit{scaled} and \textit{rounded} case).
On top of that, we show a high correlation of the average annealing result and simulated annealing for the \textit{scaled} and \textit{rounded} variant.
Therefore, it is possible to (locally) extrapolate to higher problem instances by using classical simulated annealing and then use quantum annealing to improve the results. %\lsnote{(Outlook: For what problem sizes this still holds)}

\section{Conclusion and Outlook}
\label{sec:concl}
Creating good \ac{qubo} formulations for industry use cases is not a trivial problem: encodings, slack variables, and penalty terms play a crucial role in performance. 
Formulations that limit the number of qubits, if they exist, should be prioritized.
However, in the long term, quantum computers may have enough qubits and be sufficiently robust for naive formulations to work. In this work, we have explored the formulation and solution of an industrial optimization application using quantum computing techniques. We used real anonymized data to describe the problem of allocating production capacity for a network of press shops.

We observed that quantum annealing via D-Wave hardware performs surprisingly well, even for modest-size problem instances. This approach offers great potential, but the embedding dimension scaling with the problem size is not yet ideal. Hopefully, better architectures with better connectivity will become available in the future.

Even modest-size problems require significant quantum resources to work for all algorithms and platforms we tested.
Noisy quantum hardware and incomplete connectivity seem to be the main limiting factors for scaling quantum computers to industry-level needs.
Compared to \ac{qaoa}, \ac{lrqaoa} is simpler and cheaper to implement---using orders of magnitude less quantum resources in time.
However, when deployed on IBM machines, it lags behind the classical baseline of simulated annealing---mainly due to noise.
Although the noisy nature of quantum gates is evident from the obtained solutions, they still contain problem-specific information.
Moreover, noiseless simulations return good results for small problem instances. Hence, we expect that solution quality will improve with \acp{qecc}.

Understanding the limitations of quantum algorithms for industry-relevant problems is still an open problem. 
It is important to not only hand-pick idealized problems, but to test the complexity of real data and real use cases.
Moreover, industrial adoption of quantum technologies requires building a strong software infrastructure and (automated) toolchains. Although, specialized knowledge (\ie, quantum physics) is still needed to program and make sense of results of quantum optimization, the entry barrier for practitioners is being continuously lowered by software development, which is reflected in the relative ease of use of tools such as those from vendors like D-Wave or IBM.

\newcommand{\LS}{\censor{LS}\xspace}
\newcommand{\WM}{\censor{WM}\xspace}
\newcommand{\VJ}{\censor{VJ}\xspace}
\newcommand{\USWE}{\censor{US}\xspace}
\newcommand{\THUS}{\censor{TH}\xspace}

\newcommand{\programme}{\blackout{German Federal Ministry of
    Education and Research (BMBF) funding program \enquote{Quantum Technologies---from
      Basic Research to Market}}}
\newcommand{\granttaqopam}{\censor{\#13N15647}}
\newcommand{\grantoth}{\censor{\#13N16092}}
\newcommand{\grantopt}{\censor{\#13N16095}}
\newcommand{\hta}{\censor{High-Tech Agenda Bavaria}}

\begin{small}
  \noindent\textbf{Acknowledgments} 
  This project was carried out within the TAQO-PAM consortium from the \programme, grant \granttaqopam{}, with funding codes \grantoth{} (\LS, \WM), \grantopt{} (\VJ, \USWE, \THUS) and \#13N16268 (CAR, FH). \WM acknowledges support by the \hta.
  Furthermore, we thank Martin Zehetmaier, Sebastian Ortlepp, and Carsten Tham, from BMW, and Georg Fraunhofer from OptWare GmbH for valuable discussions and proofreading the manuscript.
\end{small}

\bibliographystyle{IEEEtran}
\bibliography{references.bib}

% Generated by IEEEtran.bst, version: 1.14 (2015/08/26)
\begin{thebibliography}{10}
\providecommand{\url}[1]{#1}
\csname url@samestyle\endcsname
\providecommand{\newblock}{\relax}
\providecommand{\bibinfo}[2]{#2}
\providecommand{\BIBentrySTDinterwordspacing}{\spaceskip=0pt\relax}
\providecommand{\BIBentryALTinterwordstretchfactor}{4}
\providecommand{\BIBentryALTinterwordspacing}{\spaceskip=\fontdimen2\font plus
\BIBentryALTinterwordstretchfactor\fontdimen3\font minus \fontdimen4\font\relax}
\providecommand{\BIBforeignlanguage}[2]{{%
\expandafter\ifx\csname l@#1\endcsname\relax
\typeout{** WARNING: IEEEtran.bst: No hyphenation pattern has been}%
\typeout{** loaded for the language `#1'. Using the pattern for}%
\typeout{** the default language instead.}%
\else
\language=\csname l@#1\endcsname
\fi
#2}}
\providecommand{\BIBdecl}{\relax}
\BIBdecl

\bibitem{farhi2014quantum}
E.~Farhi, J.~Goldstone, and S.~Gutmann, ``A quantum approximate optimization algorithm,'' \emph{arXiv preprint arXiv:1411.4028}, 2014.

\bibitem{Bravyi2020}
\BIBentryALTinterwordspacing
S.~Bravyi, A.~Kliesch, R.~Koenig, and E.~Tang, ``Obstacles to variational quantum optimization from symmetry protection,'' \emph{Phys. Rev. Lett.}, vol. 125, p. 260505, Dec 2020. [Online]. Available: \url{https://link.aps.org/doi/10.1103/PhysRevLett.125.260505}
\BIBentrySTDinterwordspacing

\bibitem{Bravyi2022hybridquantum}
\BIBentryALTinterwordspacing
------, ``Hybrid quantum-classical algorithms for approximate graph coloring,'' \emph{Quantum}, vol.~6, p. 678, Mar. 2022. [Online]. Available: \url{http://dx.doi.org/10.22331/q-2022-03-30-678}
\BIBentrySTDinterwordspacing

\bibitem{montanezbarrera2024universalqaoaprotocolevidence}
\BIBentryALTinterwordspacing
J.~A. Montanez-Barrera and K.~Michielsen, ``Towards a universal qaoa protocol: Evidence of a scaling advantage in solving some combinatorial optimization problems,'' 2024. [Online]. Available: \url{https://arxiv.org/abs/2405.09169}
\BIBentrySTDinterwordspacing

\bibitem{RevModPhys.90.015002}
\BIBentryALTinterwordspacing
T.~Albash and D.~A. Lidar, ``Adiabatic quantum computation,'' \emph{Rev. Mod. Phys.}, vol.~90, p. 015002, Jan 2018. [Online]. Available: \url{https://link.aps.org/doi/10.1103/RevModPhys.90.015002}
\BIBentrySTDinterwordspacing

\bibitem{Wurtz2022counterdiabaticity}
\BIBentryALTinterwordspacing
J.~Wurtz and P.~J. Love, ``Counterdiabaticity and the quantum approximate optimization algorithm,'' \emph{{Quantum}}, vol.~6, p. 635, Jan. 2022. [Online]. Available: \url{https://doi.org/10.22331/q-2022-01-27-635}
\BIBentrySTDinterwordspacing

\bibitem{BLEKOS20241}
\BIBentryALTinterwordspacing
K.~Blekos, D.~Brand, A.~Ceschini, C.-H. Chou, R.-H. Li \emph{et~al.}, ``A review on quantum approximate optimization algorithm and its variants,'' \emph{Physics Reports}, vol. 1068, pp. 1--66, 2024, a review on Quantum Approximate Optimization Algorithm and its variants. [Online]. Available: \url{https://www.sciencedirect.com/science/article/pii/S0370157324001078}
\BIBentrySTDinterwordspacing

\bibitem{10.3389/fphy.2014.00005}
\BIBentryALTinterwordspacing
A.~Lucas, ``Ising formulations of many np problems,'' \emph{Frontiers in Physics}, vol.~2, 2014. [Online]. Available: \url{https://www.frontiersin.org/journals/physics/articles/10.3389/fphy.2014.00005}
\BIBentrySTDinterwordspacing

\bibitem{montanaro2024quantumspeedupssolvingnearsymmetric}
\BIBentryALTinterwordspacing
A.~Montanaro and L.~Zhou, ``Quantum speedups in solving near-symmetric optimization problems by low-depth qaoa,'' 2024. [Online]. Available: \url{https://arxiv.org/abs/2411.04979}
\BIBentrySTDinterwordspacing

\bibitem{doi:10.1126/sciadv.adm6761}
\BIBentryALTinterwordspacing
R.~Shaydulin, C.~Li, S.~Chakrabarti, M.~DeCross, D.~Herman \emph{et~al.}, ``Evidence of scaling advantage for the quantum approximate optimization algorithm on a classically intractable problem,'' \emph{Science Advances}, vol.~10, no.~22, p. eadm6761, 2024. [Online]. Available: \url{https://www.science.org/doi/abs/10.1126/sciadv.adm6761}
\BIBentrySTDinterwordspacing

\bibitem{sachdeva2024quantumoptimizationusing127qubit}
\BIBentryALTinterwordspacing
N.~Sachdeva, G.~S. Hartnett, S.~Maity, S.~Marsh, Y.~Wang \emph{et~al.}, ``Quantum optimization using a 127-qubit gate-model ibm quantum computer can outperform quantum annealers for nontrivial binary optimization problems,'' 2024. [Online]. Available: \url{https://arxiv.org/abs/2406.01743}
\BIBentrySTDinterwordspacing

\bibitem{preskill2018}
\BIBentryALTinterwordspacing
J.~Preskill, ``Quantum computing in the nisq era and beyond,'' \emph{Quantum}, vol.~2, p.~79, 2018. [Online]. Available: \url{https://doi.org/10.22331/q-2018-08-06-79}
\BIBentrySTDinterwordspacing

\bibitem{RevModPhys.94.015004}
\BIBentryALTinterwordspacing
K.~Bharti, A.~Cervera-Lierta, T.~H. Kyaw, T.~Haug, S.~Alperin-Lea \emph{et~al.}, ``Noisy intermediate-scale quantum algorithms,'' \emph{Rev. Mod. Phys.}, vol.~94, p. 015004, Feb 2022. [Online]. Available: \url{https://link.aps.org/doi/10.1103/RevModPhys.94.015004}
\BIBentrySTDinterwordspacing

\bibitem{10.1145/227683.227684}
\BIBentryALTinterwordspacing
M.~X. Goemans and D.~P. Williamson, ``Improved approximation algorithms for maximum cut and satisfiability problems using semidefinite programming,'' \emph{J. ACM}, vol.~42, no.~6, p. 1115–1145, Nov. 1995. [Online]. Available: \url{https://doi.org/10.1145/227683.227684}
\BIBentrySTDinterwordspacing

\bibitem{63205816-165a-3cfa-bc71-76de0e788f1a}
\BIBentryALTinterwordspacing
D.~L. Applegate, R.~E. Bixby, V.~Chvatál, and W.~J. Cook, \emph{The Traveling Salesman Problem: A Computational Study}.\hskip 1em plus 0.5em minus 0.4em\relax Princeton University Press, 2006. [Online]. Available: \url{http://www.jstor.org/stable/j.ctt7s8xg}
\BIBentrySTDinterwordspacing

\bibitem{Nenno_2024}
D.~M. Nenno and A.~Caspari, ``Dynamic optimization on quantum hardware: Feasibility for a process industry use case,'' \emph{Computers \& Chemical Engineering}, vol. 186, p. 108704, 2024.

\bibitem{Deng_2023}
\BIBentryALTinterwordspacing
Z.~Deng, X.~Wang, and B.~Dong, ``Quantum computing for future real-time building hvac controls,'' \emph{Applied Energy}, vol. 334, p. 120621, 2023. [Online]. Available: \url{https://www.sciencedirect.com/science/article/pii/S0306261922018785}
\BIBentrySTDinterwordspacing

\bibitem{fernandezvillaverde_2023}
\BIBentryALTinterwordspacing
J.~Fernández-Villaverde and I.~Hull, ``Dynamic programming on a quantum annealer: Solving the rbc model,'' 2023. [Online]. Available: \url{https://arxiv.org/abs/2306.04285}
\BIBentrySTDinterwordspacing

\bibitem{haener_2018}
\BIBentryALTinterwordspacing
T.~Häner, M.~Roetteler, and K.~M. Svore, ``Optimizing quantum circuits for arithmetic,'' 2018. [Online]. Available: \url{https://arxiv.org/abs/1805.12445}
\BIBentrySTDinterwordspacing

\bibitem{schoenberger:23:pvldb}
\BIBentryALTinterwordspacing
M.~Schönberger, I.~Trummer, and W.~Mauerer, ``\BIBforeignlanguage{en}{Quantum-inspired digital annealing for join ordering},'' in \emph{\BIBforeignlanguage{en}{Proceedings of the VLDB Endowment}}, vol.~17, no.~3, 11 2023. [Online]. Available: \url{https://doi.org/10.14778/3632093.3632112}
\BIBentrySTDinterwordspacing

\bibitem{Glover_2019}
\BIBentryALTinterwordspacing
F.~Glover, G.~Kochenberger, and Y.~Du, ``A tutorial on formulating and using qubo models,'' 2019. [Online]. Available: \url{https://arxiv.org/abs/1811.11538}
\BIBentrySTDinterwordspacing

\bibitem{Schuetz_2022}
\BIBentryALTinterwordspacing
M.~J. Schuetz, J.~K. Brubaker, H.~Montagu, Y.~van Dijk, J.~Klepsch \emph{et~al.}, ``Optimization of robot-trajectory planning with nature-inspired and hybrid quantum algorithms,'' \emph{Phys. Rev. Appl.}, vol.~18, p. 054045, Nov 2022. [Online]. Available: \url{https://link.aps.org/doi/10.1103/PhysRevApplied.18.054045}
\BIBentrySTDinterwordspacing

\bibitem{Schmidbauer_2024}
\BIBentryALTinterwordspacing
L.~Schmidbauer, K.~Wintersperger, E.~Lobe, and W.~Mauerer, ``Polynomial reduction methods and their impact on qaoa circuits,'' in \emph{2024 IEEE International Conference on Quantum Software (QSW)}, vol. 986.\hskip 1em plus 0.5em minus 0.4em\relax IEEE, Jul. 2024, p. 35–45. [Online]. Available: \url{http://dx.doi.org/10.1109/QSW62656.2024.00018}
\BIBentrySTDinterwordspacing

\bibitem{Hauke_2020}
\BIBentryALTinterwordspacing
P.~Hauke, H.~G. Katzgraber, W.~Lechner, H.~Nishimori, and W.~D. Oliver, ``Perspectives of quantum annealing: methods and implementations,'' \emph{Reports on Progress in Physics}, vol.~83, no.~5, p. 054401, may 2020. [Online]. Available: \url{https://dx.doi.org/10.1088/1361-6633/ab85b8}
\BIBentrySTDinterwordspacing

\bibitem{Yarkoni_2022}
\BIBentryALTinterwordspacing
S.~Yarkoni, E.~Raponi, T.~Bäck, and S.~Schmitt, ``Quantum annealing for industry applications: introduction and review,'' \emph{Reports on Progress in Physics}, vol.~85, no.~10, p. 104001, sep 2022. [Online]. Available: \url{https://dx.doi.org/10.1088/1361-6633/ac8c54}
\BIBentrySTDinterwordspacing

\bibitem{Vandelli2024}
\BIBentryALTinterwordspacing
M.~Vandelli, A.~Lignarolo, C.~Cavazzoni, and D.~Dragoni, ``Evaluating the practicality of quantum optimization algorithms for prototypical industrial applications,'' \emph{Quantum Information Processing}, vol.~23, no.~10, p. 344, Oct 2024. [Online]. Available: \url{https://doi.org/10.1007/s11128-024-04560-1}
\BIBentrySTDinterwordspacing

\bibitem{krol2024qissquantumindustrialshift}
\BIBentryALTinterwordspacing
A.~M. Krol, M.~Erdmann, R.~Mishra, P.~Singkanipa, E.~Munro \emph{et~al.}, ``Qiss: Quantum industrial shift scheduling algorithm,'' 2024. [Online]. Available: \url{https://arxiv.org/abs/2401.07763}
\BIBentrySTDinterwordspacing

\bibitem{Morales_2020}
\BIBentryALTinterwordspacing
M.~E.~S. Morales, J.~D. Biamonte, and Z.~Zimborás, ``On the universality of the quantum approximate optimization algorithm,'' \emph{Quantum Information Processing}, vol.~19, no.~9, Aug. 2020. [Online]. Available: \url{http://dx.doi.org/10.1007/s11128-020-02748-9}
\BIBentrySTDinterwordspacing

\bibitem{montanezbarrera_2024}
\BIBentryALTinterwordspacing
J.~A. Montanez-Barrera and K.~Michielsen, ``Towards a universal qaoa protocol: Evidence of a scaling advantage in solving some combinatorial optimization problems,'' 2024. [Online]. Available: \url{https://arxiv.org/abs/2405.09169}
\BIBentrySTDinterwordspacing

\bibitem{Aharonov_2008}
D.~Aharonov, W.~van Dam, J.~Kempe, Z.~Landau, S.~Lloyd \emph{et~al.}, ``\BIBforeignlanguage{en}{Adiabatic quantum computation is equivalent to standard quantum computation},'' \emph{\BIBforeignlanguage{en}{SIAM Rev. Soc. Ind. Appl. Math.}}, vol.~50, no.~4, pp. 755--787, Jan. 2008.

\bibitem{McGeoch_2014}
C.~C. McGeoch, \emph{Adiabatic quantum computation and quantum annealing}, ser. Synthesis lectures on quantum computing.\hskip 1em plus 0.5em minus 0.4em\relax Cham: Springer International Publishing, 2014.

\bibitem{Choi_2020}
V.~Choi, ``\BIBforeignlanguage{en}{The effects of the problem hamiltonian parameters on the minimum spectral gap in adiabatic quantum optimization},'' \emph{\BIBforeignlanguage{en}{Quantum Inf. Process.}}, vol.~19, no.~3, Mar. 2020.

\bibitem{Braida_2021}
A.~Braida and S.~Martiel, ``\BIBforeignlanguage{en}{Anti-crossings and spectral gap during quantum adiabatic evolution},'' \emph{\BIBforeignlanguage{en}{Quantum Inf. Process.}}, vol.~20, no.~8, Aug. 2021.

\bibitem{Fujii_2023}
T.~Fujii, K.~Komuro, Y.~Okudaira, and M.~Sawada, ``\BIBforeignlanguage{en}{Eigenvalue-invariant transformation of ising problem for anti-crossing mitigation in quantum annealing},'' \emph{\BIBforeignlanguage{en}{J. Phys. Soc. Jpn.}}, vol.~92, no.~4, Apr. 2023.

\bibitem{Zaborniak_2021}
\BIBentryALTinterwordspacing
T.~Zaborniak and R.~de~Sousa, ``Benchmarking hamiltonian noise in the d-wave quantum annealer,'' \emph{IEEE Transactions on Quantum Engineering}, vol.~2, p. 1–6, 2021. [Online]. Available: \url{http://dx.doi.org/10.1109/TQE.2021.3050449}
\BIBentrySTDinterwordspacing

\bibitem{imoto_2024}
\BIBentryALTinterwordspacing
T.~Imoto, Y.~Susa, R.~Miyazaki, T.~Kadowaki, and Y.~Matsuzaki, ``Universal quantum computation using quantum annealing with the transverse-field ising hamiltonian,'' 2024. [Online]. Available: \url{https://arxiv.org/abs/2402.19114}
\BIBentrySTDinterwordspacing

\bibitem{Nikolaev_2010}
A.~G. Nikolaev and S.~H. Jacobson, ``Simulated annealing,'' in \emph{International Series in Operations Research \& Management Science}, ser. International series in operations research \& management science.\hskip 1em plus 0.5em minus 0.4em\relax Boston, MA: Springer US, 2010, pp. 1--39.

\bibitem{SCIP}
\BIBentryALTinterwordspacing
K.~Bestuzheva, M.~Besan\c{c}on, W.-K. Chen, A.~Chmiela, T.~Donkiewicz \emph{et~al.}, ``Enabling research through the scip optimization suite 8.0,'' \emph{ACM Trans. Math. Softw.}, vol.~49, no.~2, jun 2023. [Online]. Available: \url{https://doi.org/10.1145/3585516}
\BIBentrySTDinterwordspacing

\bibitem{ortools}
\BIBentryALTinterwordspacing
L.~Perron and V.~Furnon, ``Or-tools,'' Google. [Online]. Available: \url{https://developers.google.com/optimization/}
\BIBentrySTDinterwordspacing

\bibitem{qiskit2024}
\BIBentryALTinterwordspacing
A.~Javadi-Abhari, M.~Treinish, K.~Krsulich, C.~J. Wood, J.~Lishman \emph{et~al.}, ``Quantum computing with {Q}iskit,'' 2024. [Online]. Available: \url{https://arxiv.org/abs/2405.08810}
\BIBentrySTDinterwordspacing

\bibitem{Broadbent_2009}
\BIBentryALTinterwordspacing
A.~Broadbent and E.~Kashefi, ``Parallelizing quantum circuits,'' \emph{Theoretical Computer Science}, vol. 410, no.~26, p. 2489–2510, Jun. 2009. [Online]. Available: \url{http://dx.doi.org/10.1016/j.tcs.2008.12.046}
\BIBentrySTDinterwordspacing

\bibitem{diestel_2005}
\BIBentryALTinterwordspacing
R.~Diestel, \emph{Graph Theory}, 3rd~ed.\hskip 1em plus 0.5em minus 0.4em\relax Springer-Verlag Heidelberg, New York, 2005. [Online]. Available: \url{http://www.math.ubc.ca/~solymosi/2007/443/GraphTheoryIII.pdf}
\BIBentrySTDinterwordspacing

\bibitem{kattemolle_2024}
\BIBentryALTinterwordspacing
J.~Kattemölle, ``Edge coloring lattice graphs,'' 2024. [Online]. Available: \url{https://arxiv.org/abs/2402.08752}
\BIBentrySTDinterwordspacing

\bibitem{Kim_2023}
Y.~Kim, A.~Eddins, S.~Anand, K.~X. Wei, E.~van~den Berg \emph{et~al.}, ``\BIBforeignlanguage{en}{Evidence for the utility of quantum computing before fault tolerance},'' \emph{\BIBforeignlanguage{en}{Nature}}, vol. 618, no. 7965, pp. 500--505, Jun. 2023.

\bibitem{qiskit_2024}
A.~Javadi-Abhari, M.~Treinish, K.~Krsulich, C.~J. Wood, J.~Lishman \emph{et~al.}, ``Quantum computing with {Q}iskit,'' 2024.

\bibitem{dattani2019pegasussecondconnectivitygraph}
\BIBentryALTinterwordspacing
N.~Dattani, S.~Szalay, and N.~Chancellor, ``Pegasus: The second connectivity graph for large-scale quantum annealing hardware,'' 2019. [Online]. Available: \url{https://arxiv.org/abs/1901.07636}
\BIBentrySTDinterwordspacing

\bibitem{cai_2014}
\BIBentryALTinterwordspacing
J.~Cai, W.~G. Macready, and A.~Roy, ``A practical heuristic for finding graph minors,'' 2014. [Online]. Available: \url{https://arxiv.org/abs/1406.2741}
\BIBentrySTDinterwordspacing

\bibitem{Dvorak2025}
Z.~Dvo{\v r}{\'a}k, \emph{\BIBforeignlanguage{en}{Graph minors}}.\hskip 1em plus 0.5em minus 0.4em\relax Cham, Switzerland: Springer International Publishing, May 2025.

\bibitem{Ball_2023}
\BIBentryALTinterwordspacing
S.~Ball, A.~Centelles, and F.~Huber, ``Quantum error-correcting codes and their geometries,'' \emph{Annales de l’Institut Henri Poincaré D, Combinatorics, Physics and their Interactions}, vol.~10, no.~2, p. 337–405, Feb. 2023. [Online]. Available: \url{http://dx.doi.org/10.4171/aihpd/160}
\BIBentrySTDinterwordspacing

\bibitem{Niven_2012}
\BIBentryALTinterwordspacing
E.~B. Niven and C.~V. Deutsch, ``Calculating a robust correlation coefficient and quantifying its uncertainty,'' \emph{Computers and Geosciences}, vol.~40, p. 1–9, Mar. 2012. [Online]. Available: \url{http://dx.doi.org/10.1016/j.cageo.2011.06.021}
\BIBentrySTDinterwordspacing

\bibitem{Shende_2006}
V.~Shende, S.~Bullock, and I.~Markov, ``Synthesis of quantum-logic circuits,'' \emph{IEEE Transactions on Computer-Aided Design of Integrated Circuits and Systems}, vol.~25, no.~6, pp. 1000--1010, 2006.

\bibitem{Nielsen_2012}
M.~A. Nielsen and I.~L. Chuang, \emph{{Quantum Computation and Quantum Information}}.\hskip 1em plus 0.5em minus 0.4em\relax Cambridge University Press, 6 2012.

\bibitem{Zbinden_2020}
\BIBentryALTinterwordspacing
S.~Zbinden, A.~B\"{a}rtschi, H.~Djidjev, and S.~Eidenbenz, \emph{Embedding Algorithms for Quantum Annealers with Chimera and Pegasus Connection Topologies}.\hskip 1em plus 0.5em minus 0.4em\relax Springer International Publishing, 2020, p. 187–206. [Online]. Available: \url{http://dx.doi.org/10.1007/978-3-030-50743-5_10}
\BIBentrySTDinterwordspacing

\end{thebibliography}

\end{document}